\documentclass{PoS}
\usepackage{bm}

\title{Enhanced production of \bm{$\Lambda_{c}$} in proton-proton collisions at the LHC}

\ShortTitle{$\Lambda_{c}$ at the LHC}

\author{\speaker{Rafa{\l} Maciu{\l}a}\thanks{This study was partially supported by the Polish National Science Center
grant DEC-2014/15/B/ST2/02528 and by the Center for Innovation and
Transfer of Natural Sciences and Engineering Knowledge in
Rzesz{\'o}w.}\\
        Institute of Nuclear
Physics, Polish Academy of Sciences, Radzikowskiego 152, PL-31-342 Krak{\'o}w, Poland\\
        E-mail: \email{rafal.maciula@ifj.edu.pl}}

\author{Antoni Szczurek\\
        Institute of Nuclear
Physics, Polish Academy of Sciences, Radzikowskiego 152, PL-31-342 Krak{\'o}w, Poland\\
        E-mail: \email{antoni.szczurek@ifj.edu.pl}}

\abstract{We calculate cross section for production of $D$ mesons and $\Lambda_c$ baryons 
in proton-proton collisions at the LHC. The cross section for 
production of $c \bar c$ pairs is calculated 
within $k_T$-factorization approach with the Kimber-Martin-Ryskin unintegrated gluon distributions.
We show that our approach well describes the $D^0$, $D^+$ and $D_s$
experimental data. We try to understand recent ALICE and LHCb data for $\Lambda_c$
production with the $c \to \Lambda_c$ independent 
parton fragmentation approach. 
The Peterson fragmentation functions are used. 
The $f_{c \to \Lambda_c}$ fragmentation fraction and 
$\varepsilon_{c}^{\Lambda}$ parameter for $c \to \Lambda_c$ are varied.
Although one can agree with the ALICE data using standard estimation of model uncertainties one
cannot describe simultaneously the ALICE and the LHCb data with the same set of parameters.
The fraction $f_{c \to \Lambda_c}$ neccessary to describe
the ALICE data is much larger than the average value obtained
from $e^+ e^-$ or $e p$ experiments.
It seems very difficult, if not impossible, to understand
the ALICE data within the considered independent parton fragmentation scheme.}

\FullConference{XXVII International Workshop on Deep-Inelastic Scattering and Related Subjects - DIS2019\\
		8-12 April, 2019\\
		Torino, Italy}

\begin{document}

\section{Introduction}

Production of charm ($c \bar c$-pairs)
belongs in principle to the domain of perturbative physics.
The corresponding cross section can be calculated in
collinear-factorization approach. Leading-order (LO) calculation is
known to give too small cross section and rather next-to-leading order (NLO)
calculation must be performed (see \textit{e.g.} Refs.~~\cite{Nason:1989zy,Beenakker:1990maa}). An effective and efficient alternative is
$k_T$-factorization approach \cite{Catani:1990xk,Catani:1990eg,Collins:1991ty}. The $k_T$-factorization
provides a good description of $D$ meson production cross sections at 
RHIC \cite{Maciula:2015kea}, Tevatron \cite{Jung:2010ey} 
and at the LHC \cite{Maciula:2013wg,Karpishkov:2016hnx}. 

The production of $D$ mesons and/or nonphotonic leptons
requires a nonperturbative information about hadronization process.
To describe $D$ meson production fragmentation
functions (FFs) for $c \to D$ quark-to-meson transitions are usually 
included. In the context of heavy-flavour production the Peterson FFs \cite{Peterson:1982ak} are usually used. 

Recently the LHCb \cite{Aaij:2013mga} and very recently ALICE
\cite{Acharya:2017kfy} Collaborations obtained new results 
for $\Lambda_c$ production at the highest so far collision 
energy $\sqrt{s}$ = 7 TeV.
We wish to study whether the new LHCb and ALICE data can be described
consistently within the chosen scheme of 
calculation based on $c \to \Lambda_c$ fragmentation.
If yes, it would be interesting whether the $f_{c \to \Lambda_c}$
fragmentation fraction is consistent with those found in previous 
studies of $e^+ e^-$, $e p$ and $B$ meson decays.

\section{A sketch of the theoretical formalism}

\subsection{Parton-level calculations}

In the partonic part of our numerical calculations we follow the $k_{T}$-factorization approach. This approach is commonly known to be very efficient
not only for inclusive particle distributions but also for studies of kinematical correlations.
According to this approach, the transverse momenta $k_{t}$'s (virtualities) of both partons entering the hard process are taken into account and the sum of transverse momenta of the final $c$ and $\bar c$ no longer cancels. Then the differential cross section at the tree-level for the $c \bar c$-pair production reads:
\begin{eqnarray}\label{LO_kt-factorization} 
\frac{d \sigma(p p \to c \bar c \, X)}{d y_1 d y_2 d^2p_{1,t} d^2p_{2,t}} &=&
\int \frac{d^2 k_{1,t}}{\pi} \frac{d^2 k_{2,t}}{\pi}
\frac{1}{16 \pi^2 (x_1 x_2 s)^2} \; \overline{ | {\cal M}^{\mathrm{off-shell}}_{g^* g^* \to c \bar c} |^2}
 \\  
&& \times  \; \delta^{2} \left( \vec{k}_{1,t} + \vec{k}_{2,t} 
                 - \vec{p}_{1,t} - \vec{p}_{2,t} \right) \;
{\cal F}_g(x_1,k_{1,t}^2) \; {\cal F}_g(x_2,k_{2,t}^2) \; \nonumber ,   
\end{eqnarray}
where ${\cal F}_g(x_1,k_{1,t}^2)$ and ${\cal F}_g(x_2,k_{2,t}^2)$
are the unintegrated gluon distribution functions (UGDFs) for both colliding hadrons and ${\cal M}^{\mathrm{off-shell}}_{g^* g^* \to c \bar c}$ is the off-shell matrix element for the hard subprocess. The extra integration is over transverse momenta of the initial
partons. We keep exact kinematics from the very beginning and additional hard dynamics coming from transverse momenta of incident partons. Explicit treatment of the transverse part of momenta makes the approach very efficient in studies of correlation observables. The two-dimensional Dirac delta function assures momentum conservation.
The unintegrated (transverse momentum dependent) gluon distributions must be evaluated at:
\begin{equation}
x_1 = \frac{m_{1,t}}{\sqrt{s}}\exp( y_1) 
     + \frac{m_{2,t}}{\sqrt{s}}\exp( y_2), \;\;\;\;\;\;
x_2 = \frac{m_{1,t}}{\sqrt{s}}\exp(-y_1) 
     + \frac{m_{2,t}}{\sqrt{s}}\exp(-y_2), \nonumber
\end{equation}
where $m_{i,t} = \sqrt{p_{i,t}^2 + m_c^2}$ is the quark/antiquark transverse mass. In the case of charm quark production at the LHC energies, especially in the forward rapidity region, one tests very small gluon longitudinal momentum fractions $x < 10^{-5}$.  

The leading-order matrix element squared $gg \to c\bar c$ for off-shell gluons is taken here in the analytic form proposed by Catani, Ciafaloni and Hautmann (CCH) \cite{Catani:1990eg}. The calculation of higher-order corrections in the $k_t$-factorization is much more complicated than in the case of collinear approximation. However, the common statement is that actually in the $k_{t}$-factorization approach with tree-level off-shell matrix elements some part of real higher-order corrections is effectively included. This is due to possible emission of extra soft (and even hard) gluons encoded
in the unintegrated gluon densities. More details of the theoretical formalism adopted here can be found in Ref.~\cite{Maciula:2013wg}. 
  
In the numerical calculation below we have applied the Kimber-Martin-Ryskin (KMR) UGDF that is derived from a modified DGLAP-BFKL evolution equation \cite{Watt:2003mx} and has been found recently to work very well in the case of charm production at the LHC \cite{Maciula:2013wg}.
As discussed also in Ref.~\cite{Maciula:2016kkx} the $k_T$-factorization approach with the KMR UGDF gives results well consistent with collinear NLO approach.
For the calculation of the KMR distribution we used here up-to-date collinear MMHT2014 gluon PDFs \cite{Harland-Lang:2014zoa}.
The renormalization and factorization scales $\mu^2 = \mu_{R}^{2} =
\mu_{F}^{2} = \frac{m^{2}_{1,t} + m^{2}_{2,t}}{2}$ and charm quark mass 
$m_{c} = 1.5$ GeV are used in the present study. The uncertainties related to the choice of these parameters and to the collinear gluon PDFs will be discussed shortly when presenting numerical results.

\subsection{From quarks to hadrons}

According to the often used independent parton fragmentation picture, the inclusive distributions of charmed hadrons $h =D, \Lambda_c$ are obtained through a convolution of inclusive distributions of charm quarks/antiquarks and $c \to h$ fragmentation functions:
\begin{equation}
\frac{d \sigma(pp \rightarrow h X)}{d y_h d^2 p_{t,h}} \approx
\int_0^1 \frac{dz}{z^2} D_{c \to h}(z)
\frac{d \sigma(pp \rightarrow c X)}{d y_c d^2 p_{t,c}}
\Bigg\vert_{y_c = y_h \atop p_{t,c} = p_{t,h}/z} \;,
\label{Q_to_h}
\end{equation}
where $p_{t,c} = \frac{p_{t,h}}{z}$ and $z$ is the fraction of
longitudinal momentum of charm quark $c$ carried by a hadron $h =D, \Lambda_c$.
A typical approximation in this formalism assumes that $y_{c}$ is
unchanged in the fragmentation process, i.e. $y_h = y_c$. It was originally motivated for light hadrons
but is commonly accepted also in the case of heavy hadrons.

As a default set in all the following numerical calculations the
standard Peterson model of fragmentation function \cite{Peterson:1982ak}
with the parameters $\varepsilon_{c}^{D} = \varepsilon_{c}^{\Lambda} = 0.05$ is applied. The parameter will be varied only in the case of $c \to \Lambda_c$ transition. This choice of fragmentation function and parameters is based on our previous theoretical studies of open charm production at the LHC \cite{Maciula:2013wg}, where detailed analysis of uncertainties related to application of different models of FFs was done.
 
Finally, the calculated cross sections for $D^{0}, D^{+}, D^{+}_{S}$ mesons and $\Lambda_c$ baryon should be normalized to the relevant fragmentation fractions.
For a nice review of the charm fragmentation fractions see Ref.~\cite{Lisovyi:2015uqa}.

\section{Numerical results}

\begin{figure}[!h]
\begin{minipage}{0.47\textwidth}
 \centerline{\includegraphics[width=1.0\textwidth]{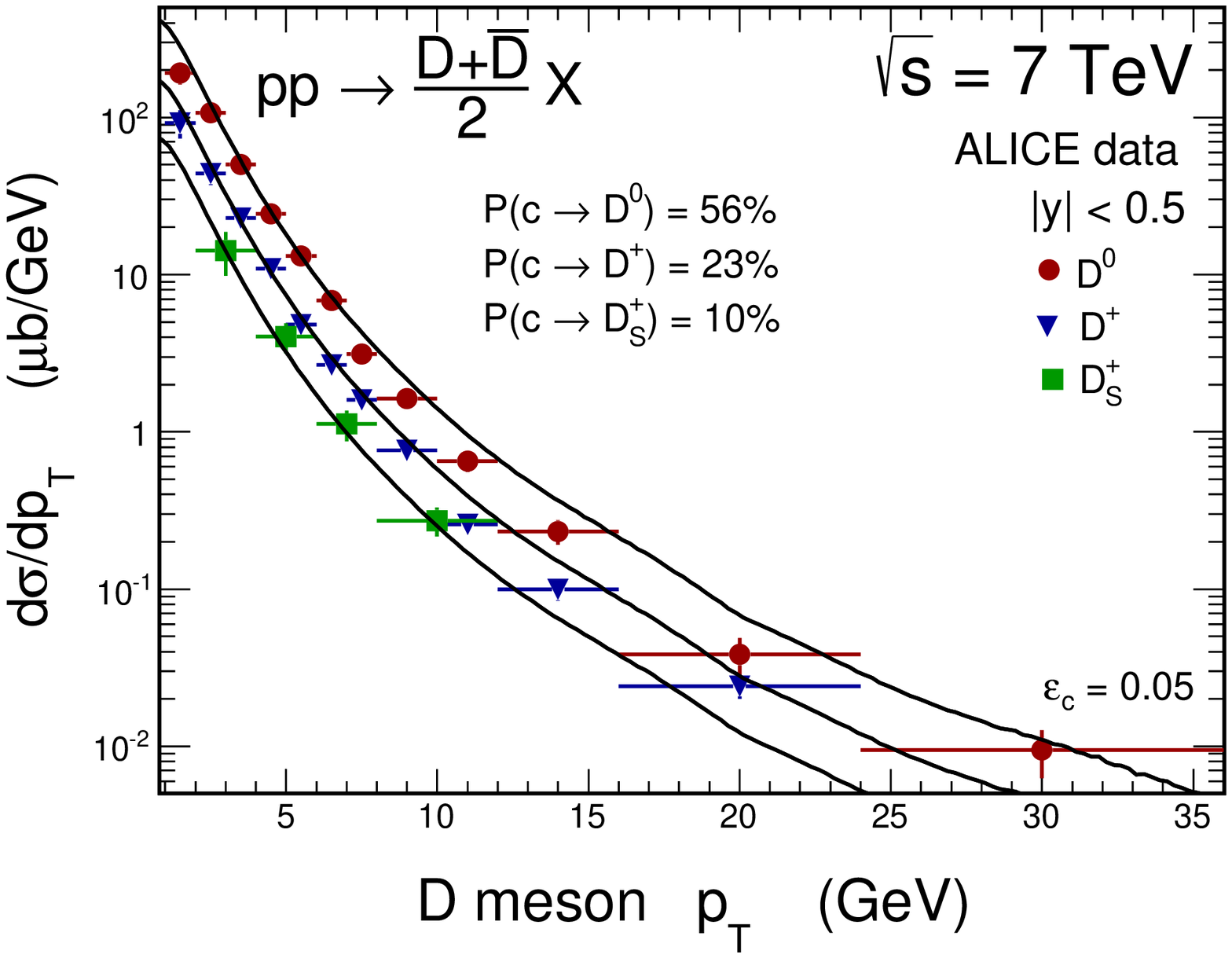}}
\end{minipage}
\begin{minipage}{0.47\textwidth}
 \centerline{\includegraphics[width=1.0\textwidth]{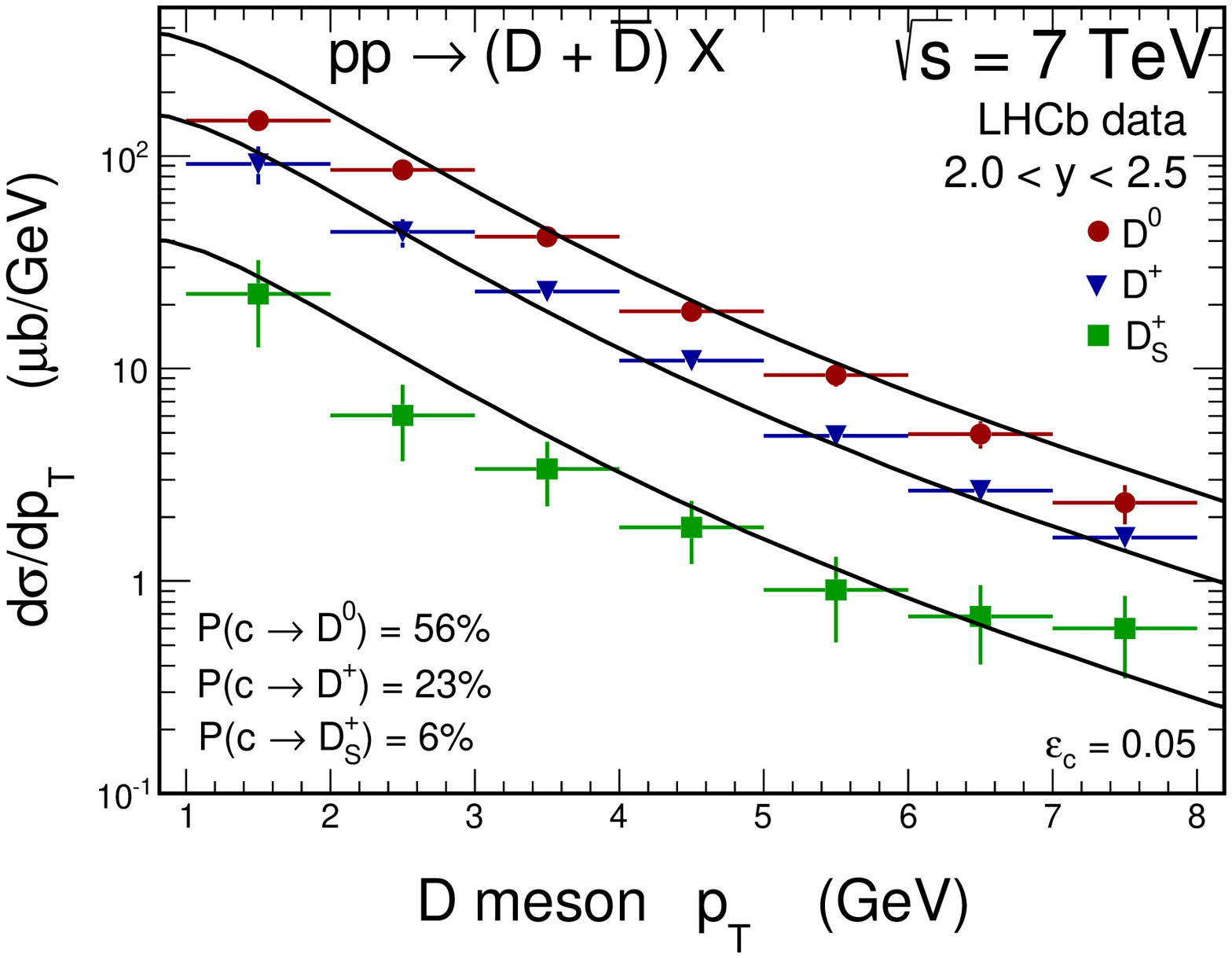}}
\end{minipage}
\caption{
\small Transverse momentum distribution of $D$ mesons for $\sqrt{s}$ = 7
TeV for ALICE (left panel) and LHCb (right panel). 
The experimental data points are taken from Refs.~\cite{Acharya:2017jgo} 
and \cite{Aaij:2013mga}, respectively.
}
 \label{fig:dsig_dpt_Dmesons}
\end{figure}

We start our presentation by showing results for $D$ meson
production. In Fig.~\ref{fig:dsig_dpt_Dmesons} we present transverse
momentum distributions of different open charm mesons - $D^0, D^+$, 
and $D_s$ for the ALICE (left panel) and the LHCb (right panel) kinematics.
Here, and throughout this subsection, the numerical results are obtained
within the standard fragmentation procedure with the assumption of
unchanged rapidity, \textit{i.e.} $y_{c} = y_{h}$, where $h=D,\Lambda_c$. In this calculation
we use standard Peterson fragmentation function with 
$\varepsilon_{c}^{D}= 0.05$ for $c \to D$ transition. 
The fragmentation fractions for charmed mesons are set to be 
$f_{c \to D^0}$ = 0.56 and $f_{c\to D^+}$ = 0.23 for both, ALICE 
and LHCb detector acceptance.
In the case of charmed-strange meson two different values of 
the fragmentation fraction are needed to fit both data sets with 
the same precision, \textit{i.e.} $f_{c \to D_S} = 0.06$ for LHCb 
and $0.10$ for ALICE. Both values of the fragmentation fraction for 
$c \to D_S$ transition are consistent with those extracted from 
combined analysis of charm-quark fragmentation fraction measurements 
in $e^+e^-$, $ep$, and $pp$ collisions \cite{Lisovyi:2015uqa}.
We cannot describe both sets of data with the same $f_{c \to D_S}$.
Doing so we would get clear disagreement using, \textit{e.g.} $\chi^2$-criterion.
It looks there is a similar effect as for $\Lambda_c$, to be discussed below.

\begin{figure}[!htbp]
\begin{minipage}{0.47\textwidth}
 \centerline{\includegraphics[width=1.0\textwidth]{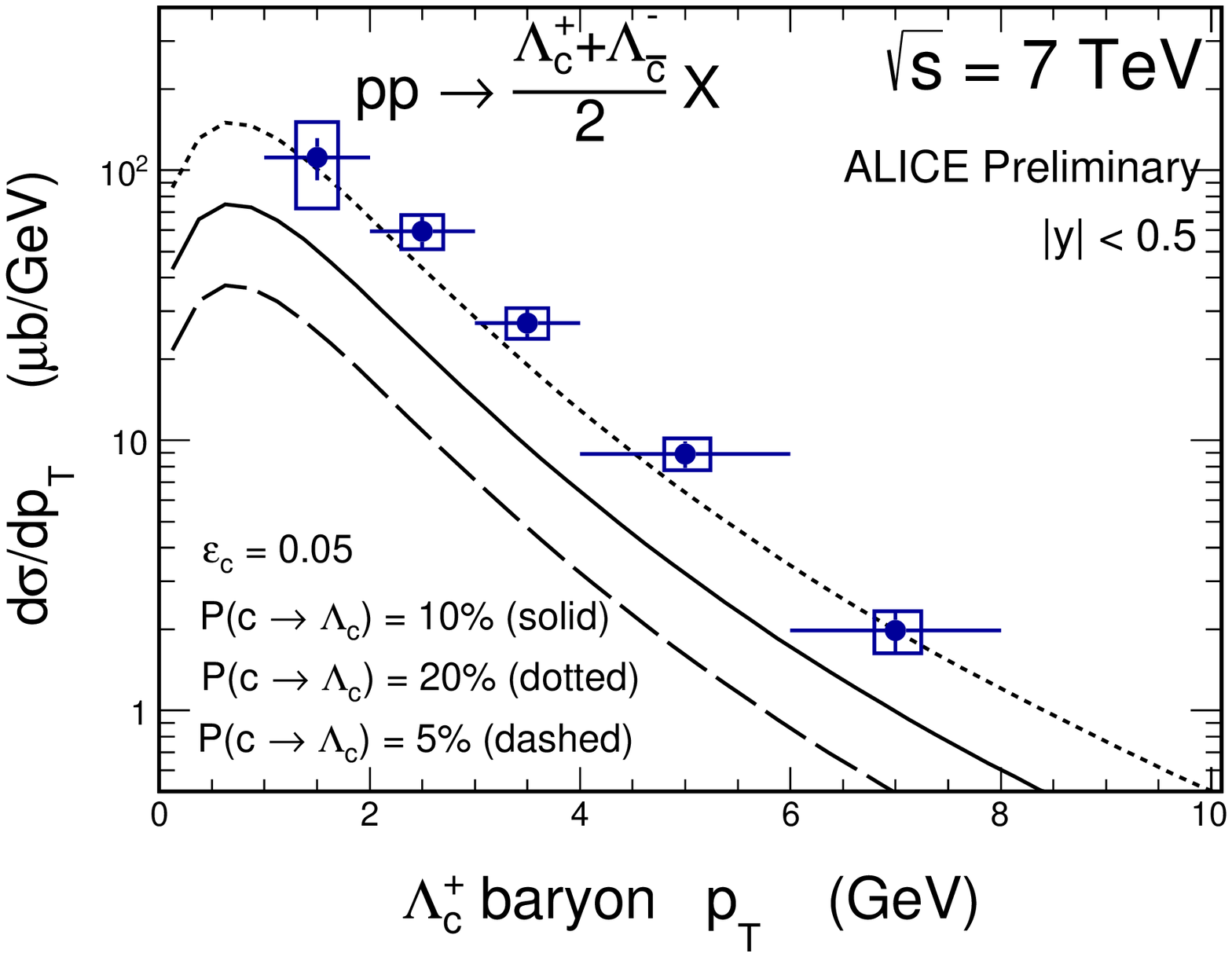}}
\end{minipage}
\begin{minipage}{0.47\textwidth}
 \centerline{\includegraphics[width=1.0\textwidth]{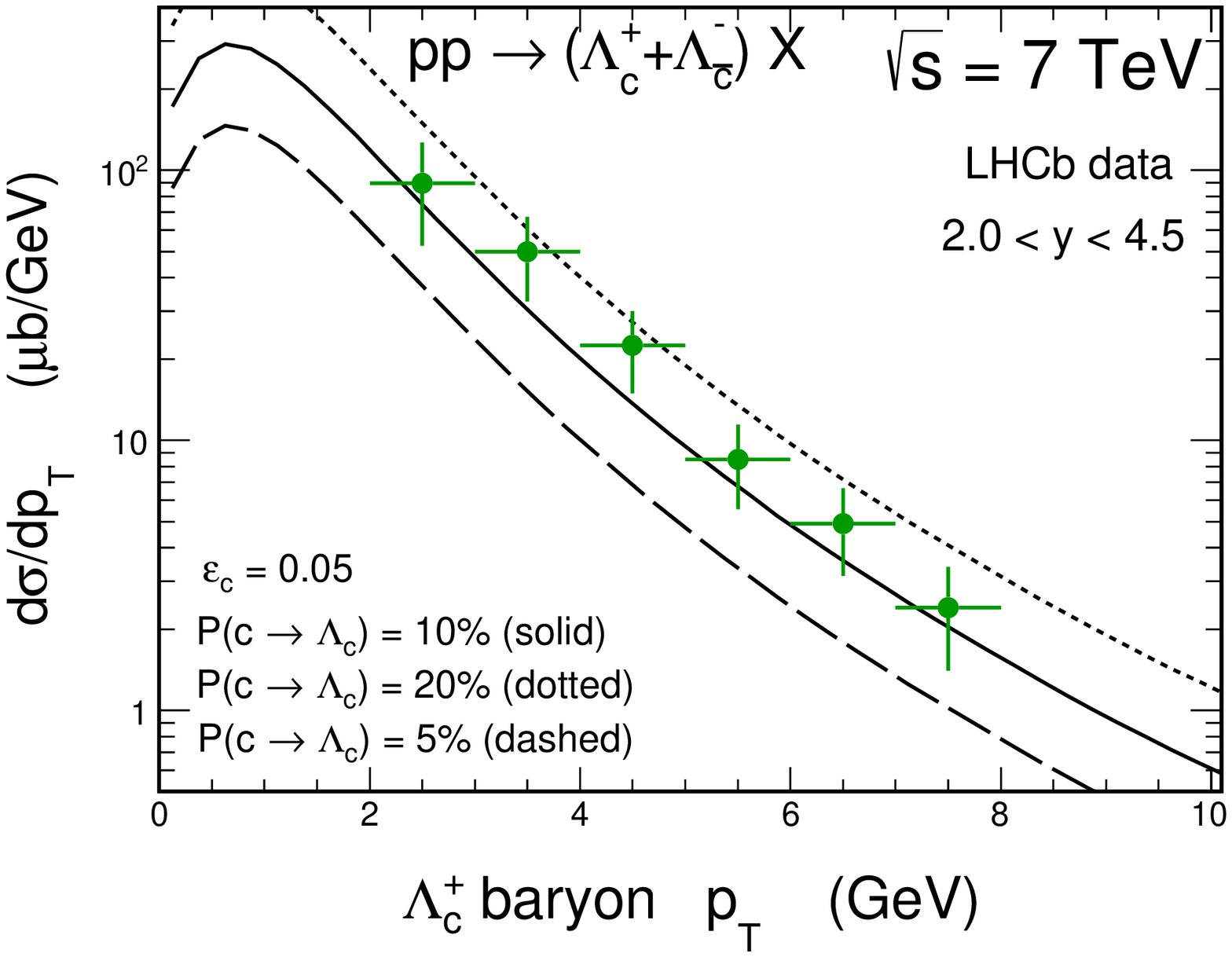}}
\end{minipage}
\caption{
\small Transverse momentum distribution of $\Lambda_c$ baryon 
for $\sqrt{s}$ = 7 TeV for ALICE (left panel) and LHCb (right panel).
The experimental data points are taken from Refs.~\cite{Acharya:2017kfy} 
and \cite{Aaij:2013mga}, respectively.
}
\label{fig:dsig_dpt_LambdaC}
\end{figure}

Having fixed all parameters of the theoretical approach in the context 
of open charm meson production we can proceed to the production 
of $\Lambda_c$ baryons. 
In Fig.~\ref{fig:dsig_dpt_LambdaC} we present transverse momentum
distribution of $\Lambda_c$ baryons for the ALICE (left panel) and 
the LHCb (right panel) kinematics. 
In this calculation we have also used the Peterson FF with the same
parameter $\varepsilon_{c}^{\Lambda} = 0.05$ (as a default) as 
for $c \to D$ transition. The three lines correspond to different values of 
$c \to \Lambda_c$ fragmentation fractions.
The dashed curve is for $f_{c \to \Lambda_c}$ = 0.05, as typical for
pre-LHC results. Clearly this result underpredicts both ALICE and LHCb
data. We show also result for increased fragmentation fractions, 
\textit{i.e.} $f_{c \to \Lambda_c}$ = 0.10 (solid line) and 0.20 (dotted line).
The agreement between data and the theory predictions with the increased
$f_{c \to \Lambda_c}$ becomes better. However, a visible difference
appears in the observed agreement for the mid-rapidity ALICE and 
for forward LHCb regimes. Taking $f_{c \to \Lambda_c}$ = 0.10 
we are able to describe the LHCb data quite well 
but we still underestimate the ALICE data by a factor $\sim 2$ in 
the whole considered range of transverse momenta. The shapes of the
transverse momentum distributions are well reproduced in both ALICE and LHCb cases.     
In order to get right normalization in the case of the ALICE measurement
we need to take $f_{c \to \Lambda_c}$ = 0.20 which is much bigger than
the numbers found in previous studies (see \textit{e.g.} a review 
in Ref.~\cite{Lisovyi:2015uqa}).

\section{Conclusions}

We find that the fragmentation fraction 
$f_{c \to \Lambda_c}$ = 0.1 - 0.15 describes the recent data
of the LHCb collaboration but fails to describe the new ALICE data.
Even for LHCb this number is slightly bigger than the values from 
the compilation of world results \cite{Lisovyi:2015uqa} obtained 
from experimental data on 
$e^+ e^-$ and $e p$ and $B$ meson decays.
Although we could agree with the ALICE data using standard estimation of model uncertainties related to factorization/renormalization scale, quark mass and PDF we were not able
to describe simultaneously the ALICE and the LHCb $\Lambda_{c}$-baryon data as well as data on $D$-meson production with the same set of parameters.

The interpretation of the increased fragmentation fraction
$c \to \Lambda_c$ is at present not clear and requires further studies, 
both on the theoretical and experimental side.

The independent parton fragmentation approach is only a simplification
which has no firm and fundamental grounds and requires tests 
to be valid approach.
At low energies an asymmetry in production of $\Lambda_c^+$ and 
$\Lambda_c^-$ was observed \cite{Lambdac_asymmetry}.
This may be related to the charm meson cloud in the nucleon \cite{Cazaroto:2013wy}
and/or recombination with proton remnants \cite{recombination}.
At high-energy this mechanism is active at large $x_F$ (or $\eta$,
probably for pseudorapidities larger than available for LHCb).
Certainly a study of $\Lambda_c^+/\Lambda_c^-$ asymmetry in LHC RunII
would be a valueable supplement. This would allow to verify the 
$c \to \Lambda_c$
``independent'' parton hadronization picture.
The new data of the ALICE Collaboration suggests a much bigger
$f_{c \to \Lambda_c}$ hadronization fraction than those obtained
in other processes and LHCb.
In principle, it could be even a creation of $\Lambda_c$ in the
quark-gluon plasma due to coalescence mechanism (see \textit{e.g.} Ref.~\cite{SLS2018}).
Such an enhancement was observed in p-Pb and Pb-Pb collisions and 
interpreted in terms of quark combination/coalescence approach 
in \cite{LSSW2017} (for p-Pb) and \cite{PMDG2017} (for Pb-Pb).
Even in the "independent" parton picture the hadronization fractions
$f_{c \to D_i}$ or $f_{c \to \Lambda_c}$ do not need to be universal and
may depend on partonic sourounding associated with the collision 
which may be, in principle, reaction and energy dependent. 
Therefore precise measurements at the LHC will allow to verify 
the picture and better understand the hadronization mechanism.

To explore experimentally the hypothesis that $\Lambda_c$ is produced in the mini quark-gluon plasma
one could study its production rates as a function of event multiplicity and compare to
similar analysis for the production of $D^{0}$ mesons.

\end{document}